\documentclass[12pt]{article}
\usepackage{geometry}
\usepackage{subfigure}
\usepackage{physics}
\usepackage{multirow}
\usepackage{booktabs}
\numberwithin{equation}{section}
\makeatletter
\newcommand\ackname{Acknowledgements}
\if@titlepage
\begin{figure}
    \centering
    \includegraphics[width=0.5\linewidth]{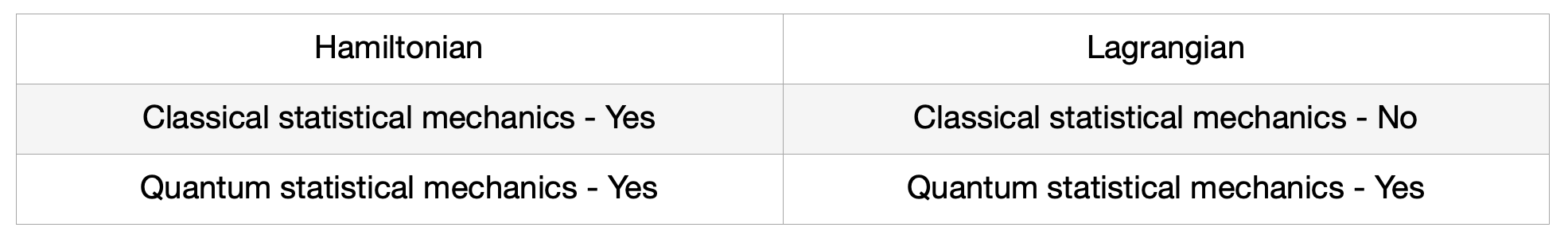}
    \caption{Enter Caption}
    \label{fig:enter-label}
\end{figure}
\newenvironment{acknowledgements}{
	\titlepage
	\null\vfil
	\@beginparpenalty\@lowpenalty
	\begin{center}
		\bfseries \ackname
		\@endparpenalty\@M1
\end{center}}
{\par\vfil\null\endtitlepage}
\else
\newenvironment{acknowledgements}{
	\if@twocolumn
	\section*{\abstractname}
	\else
	\small
	\begin{center}
		{\bfseries \ackname\vspace{-.5em}\vspace{\z@}}
	\end{center}
	\quotation
	\fi}
{\if@twocolumn\else\endquotation\fi}
\usepackage{mathrsfs}
\usepackage{eucal}
\usepackage{amssymb,amsmath}
\usepackage{pgf}
\usepackage{tikz}
\usepackage{cite}
\usepackage{graphicx}
\usepackage{color}
\usepackage{amssymb}
\usepgflibrary{arrows}
\usetikzlibrary{calc,angles,positioning,intersections,quotes,backgrounds,decorations.pathreplacing}
\usetikzlibrary{decorations.markings}
\usepackage{amsmath}
\usepackage{amsthm}
\usepackage{accents}

\theoremstyle{remark}
\theoremstyle{definition}

\usepackage[makeroom]{cancel}

\usepackage{comment}
\begin{document}
	\title{\textbf{ Lagrangian formalism and classical statistical ensemble}}
	\author{Sikarin Yoo-Kong \\	
		\small  \emph{The Institute for Fundamental Study (IF),} \\
		\small\emph{Naresuan University (NU), Phitsanulok-Nakhon Sawan,}\\
		\small  \emph{99 Moo 9, Tha Pho, Mueang Phitsanulok, 65000 Phitsanulok, Thailand.}	\\
        \small\emph{e-mail:\;sikariny@nu.ac.th}
	}
	\date{}
	\maketitle
	\abstract
	We present a formulation of classical statistical mechanics based on a Lagrangian description on the tangent bundle. In this approach, a Wick rotation from real time to imaginary time is employed as a technical device that facilitates the construction of a Hamiltonian structure expressed in velocity variables. The resulting dynamics preserves a natural measure induced by the associated symplectic form on the tangent bundle. This measure-preserving property enables the consistent definition of classical statistical ensembles directly in terms of Lagrangian variables.
	
\section{Introduction}\label{S1}
\begin{figure}[h]
\centering
\includegraphics[width=15cm]{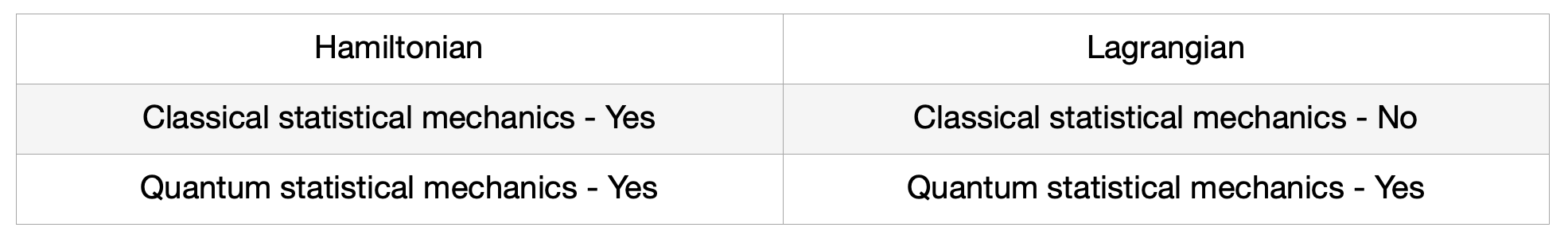}
\caption{Hamiltonian and Lagrangian approaches in the classical and quantum statistical mechanics.}\label{F1}
\end{figure}

Classical statistical mechanics is traditionally formulated within the Hamiltonian framework, where the phase space is identified with the cotangent bundle and the dynamics is governed by Hamilton’s equations \cite{FD}. In this setting, the symplectic structure and Liouville’s theorem play a central role in defining invariant measures and statistical ensembles. While this formulation is well established and highly successful, it relies fundamentally on canonical momenta and phase-space variables. On the other hand, classical mechanics itself admits an equally fundamental description in terms of the Lagrangian formalism, which is naturally defined on the tangent bundle of the configuration space. Despite its central role in analytical mechanics, the Lagrangian framework is far less explored in the context of classical statistical mechanics, see Fig \ref{F1}. This raises a natural question: to what extent can statistical ensembles be constructed directly from Lagrangian variables, without explicit reference to canonical momenta? Several attempts have been made to relate Lagrangian mechanics to statistical descriptions, often through Legendre transformations or path-integral methods. However, a direct and systematic construction of classical statistical ensembles formulated on the tangent bundle remains relatively underdeveloped. One of the main obstacles is that the standard second-order Lagrangian variational principle does not naturally provide the geometric structures, such as a symplectic form and invariant measure, that underlie statistical mechanics in phase space \cite{RF}.
\\
\\
In this work, we address this issue by formulating classical statistical mechanics on the tangent bundle using a first-order variational principle written in velocity variables. A Wick rotation from real time to imaginary time\footnote{In the quantum case, the Wick rotation is applied to the propagator in path integrals \cite{FR}.} is employed as a technical device that allows the Lagrangian to induce a positive-definite Hamiltonian structure on the tangent bundle for a class of natural Lagrangians with quadratic kinetic terms. The resulting formulation admits a symplectic structure derived directly from the action, ensuring measure-preserving dynamics in the standard Hamiltonian sense. Within this framework, classical statistical ensembles can be constructed consistently on the tangent bundle without introducing canonical momenta explicitly. The microcanonical and canonical ensembles follow from the invariant measure associated with the induced symplectic structure, and thermodynamic quantities such as entropy and partition functions can be defined in the usual way. Importantly, the resulting statistical mechanics is thermodynamically equivalent to the conventional Hamiltonian formulation for the class of systems considered. To illustrate the construction concretely, we analyze the one-dimensional harmonic oscillator as a representative example. We show that the imaginary-time Lagrangian formulation reproduces exactly the same entropy and canonical partition function as the standard Hamiltonian approach. This example serves to demonstrate the internal consistency of the framework and clarifies the relation between the tangent-bundle formulation and the familiar phase-space picture.
\\
\\
The paper is organized as follows. In Section 2, we introduce the first-order action on the tangent bundle and derive the corresponding equations of motion and symplectic structure. Section 3 is devoted to the construction of classical statistical ensembles within this framework. In Section 4, the harmonic oscillator is discussed as an explicit example. Finally, Section 5 summarizes our results and outlines possible directions for future work.
\section{Imaginary-time Lagrangian mechanics and the variational principle}\label{SS1}
In this section, we present the dynamical framework underlying our construction of statistical ensembles on the tangent bundle. The key idea is to formulate the dynamics using a first-order variational principle written in velocity variables, from which both the equations of motion and the geometric structure follow naturally. We shall first consider the Wick rotation of the Lagrangian and the variational principle. For simplicity, we shall consider a system with one degree of freedom and the Lagrangian is given by
\begin{equation}
    L(\dot q,q)=\frac{\dot q^2}{2}-V(q)\;.
\end{equation}
It is not difficult to see that, with a standard Euler-Lagrange equation, 
\begin{equation}
    \frac{\partial L}{\partial q}-\frac{d}{dt}\frac{\partial L}{\partial \dot q}=0\;,
\end{equation}
one obtains
\begin{equation}
    \ddot q=-\frac{dV}{dq}\;,
\end{equation}
which is the equation of motion.
Under the Wick rotation on time variable: $t\rightarrow i\tau$, the Lagrangian becomes
\begin{equation}
    -\mathcal L(\Tilde {q},q)=\frac{\Tilde{q}^2}{2}+V(q)\equiv E.\label{NL}
\end{equation}
where $\Tilde{q}=dq/d\tau$. We note that a new Lagrangian in \eqref{NL} can be expressed as $\mathcal{L}(\Tilde {q},q)=L(i\Tilde {q},q)$. However, we prefer to work with the notion $\mathcal{L}(\Tilde {q},q)$ as our convenient and a reason on the invariant of physics \footnote{Otherwise, we have to deal with an imaginary generalised velocity $i\Tilde{q}$, presented in the Lagrangian $L(i\Tilde{q},q)$. We, therefore, can define a new variable $\bar q=i\Tilde{q}$ which effectively results the same with a choice $\Tilde{q}$. } (which will be shortly seen later). Interestingly, 
we see that a negative Lagrangian is now a total energy of the system. With \eqref{NL}, it is not difficult to see that
\begin{equation}
    \Tilde{q}=-\frac{\partial\mathcal L}{\partial \Tilde{q}}\;,\label{q11}
\end{equation}
and the Euler-Lagrange equation with imaginary time is 
\begin{equation}
    \frac{\partial \mathcal L}{\partial q}+\frac{d}{d\tau}\frac{\partial \mathcal L}{\partial \Tilde{q}}=0\;,\label{q12}
\end{equation}
resulting in the equation of motion
\begin{equation}
    \Tilde{\Tilde{q}}=-\frac{dV}{dq}\;,
\end{equation}
which is invariant under the Wick rotation\footnote{Here, we demand the equation of motion remaining the same. Therefore, the modification of the Euler-Lagrange equation is needed resulting in \eqref{q12}.}. Inserting \eqref{q11} into \eqref{q12}, one obtains
\begin{equation}
     \Tilde{\Tilde{q}}=\frac{\partial \mathcal L}{\partial q}\;.\label{q13}
\end{equation}
Of course, the Euler-Lagrange equation \eqref{q12} could not be derived from the variational principle of the standard action functional. Therefore, we shall introduce a modified action given by
\begin{equation}
    S[q,v]=\int_{\tau'}^{\tau''}d\tau [v\tilde q+\mathcal L(q,v)]\;,
\end{equation}
where $v$ is an auxiliary variable which is treated as independent of $\tilde q$ at the level of the action. Next, we shall consider the variations $v\rightarrow v+\delta v$ and $ q\rightarrow  q+\delta  q$, resulting in
\begin{equation}
    S[ q+\delta  q, v+\delta v]=S[q,v]+\int_{\tau'}^{\tau''}d\tau \left[v\delta\tilde q+\tilde q\delta v+\frac{\partial \mathcal L}{\partial v}\delta v+\frac{\partial \mathcal L}{\partial q}\delta q\right]\;.
\end{equation}
The variation of the action is 
\begin{equation}
    \delta S=S[ q+\delta  q, v+\delta v]-S[q,v]=\int_{\tau'}^{\tau''}d\tau \left[-\tilde v\delta q+\tilde q\delta v+\frac{\partial \mathcal L}{\partial v}\delta v+\frac{\partial \mathcal L}{\partial q}\delta q\right]\;,
\end{equation}
where the integrating by parts and boundary conditions $\delta q(\tau')=\delta q(\tau'')=0$ are imposed. Rearranging things a bit, we obtain
\begin{equation}
    \delta S=\int_{\tau'}^{\tau''}d\tau \left[\left(-\tilde v+\frac{\partial \mathcal L}{\partial q}\right)\delta q+\left(\tilde q+\frac{\partial \mathcal L}{\partial v}\right)\delta v\right]\;.
\end{equation}
The critical condition of the action $\delta S=0$ gives
\begin{equation}
    \tilde v=\frac{\partial \mathcal L}{\partial q}\;,\;\;\;\tilde q=-\frac{\partial \mathcal L}{\partial \tilde q}\;.
\end{equation}
The variables $q$ and $v$ are independent off shell. However, for the class of Lagrangians considered here, the first equation in (2.4) implies a relation between $v$ and the Wick-rotated velocity $\tilde q$. This relation holds only for solutions of the equations of motion and therefore constitutes an on-shell identification $v=\tilde q$. Consequently, \eqref{q11}, \eqref{q12} and \eqref{q13} are recovered. We note here that the symplectic structure in \eqref{SSS} can be directly obtained from the action functional. To see this, we introduce the 1-form object $\theta=vdq$ and the exterior derivation $\omega=d\theta=dv\wedge dq$ is a sympletic 2-form. Imposing again the on-shell condition $v=\tilde q$, we obtain $\omega=d\theta=d\tilde q\wedge dq$ which is invariant under time evolution (see the explicit calculation in the next section).
\\
\\
\textbf{Remark}: Here, we would like to point out that although the Lagrangian \eqref{NL} becomes the total energy of the system, but this Lagrangian must be distinguished from the Hamiltonian, which is also the total energy as well, as follows. The Hamiltonian $H(p,x)$ is a function of the canonical momentum and the generalised coordinate, while the Lagrangian $L(i\Tilde{q},q)$ is a function of the imaginary generalised velocity and the generalised coordinate. This results that, with the new Lagrangian \eqref{NL}, the dynamics of the system will be expressed on the tangent bundle $(i\Tilde{q},q)$ or $(\bar q,q)$ equipped with a symplectic structure, see right below.
\section{Classical statistical ensemble}\label{SS2}
\begin{figure}
    \centering
    \includegraphics[width=0.3\linewidth]{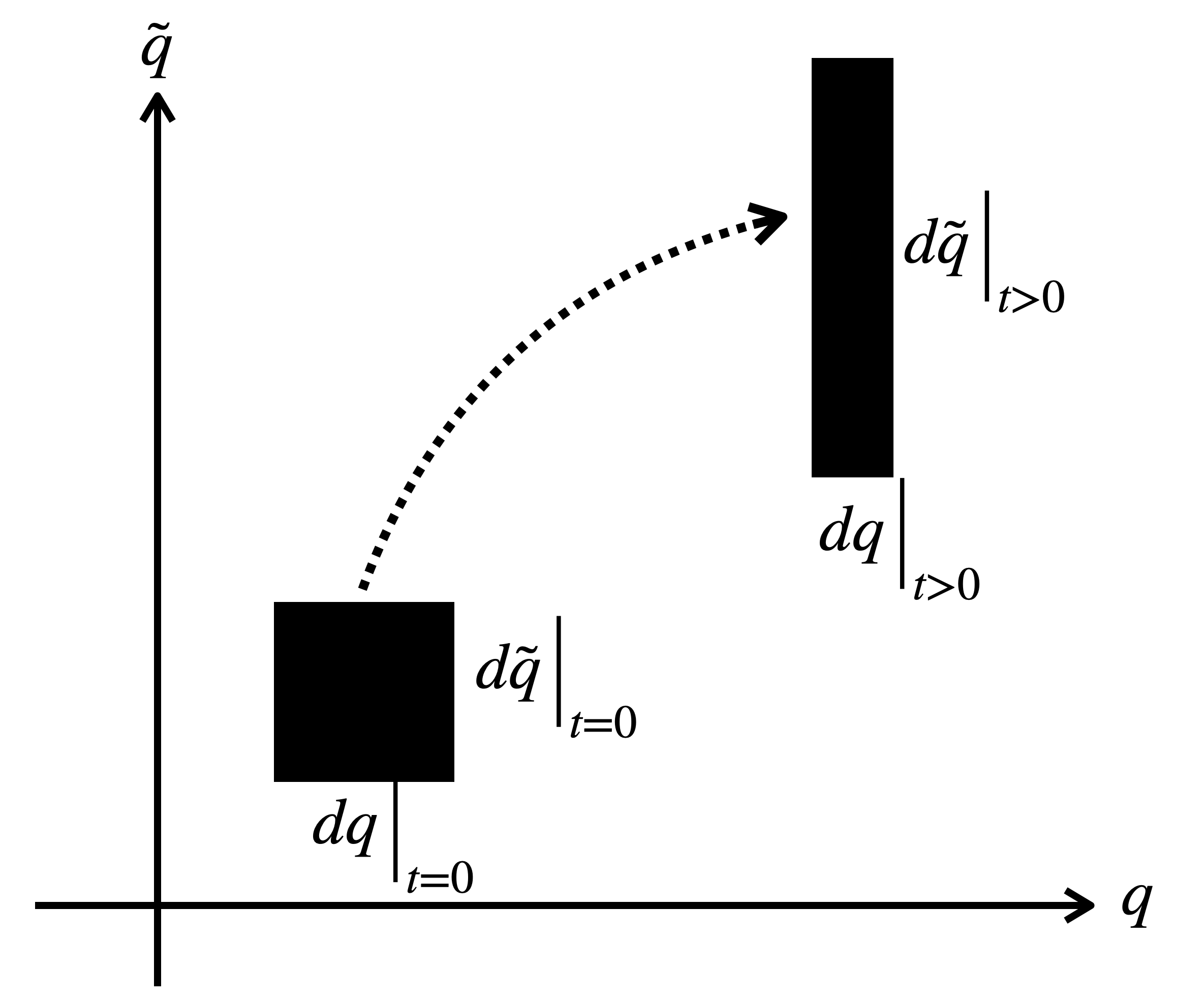}
    \caption{Area under the time evolution on the tangent bundle.}
    \label{LT1}
\end{figure}

\begin{figure}
    \centering
    \includegraphics[width=0.3\linewidth]{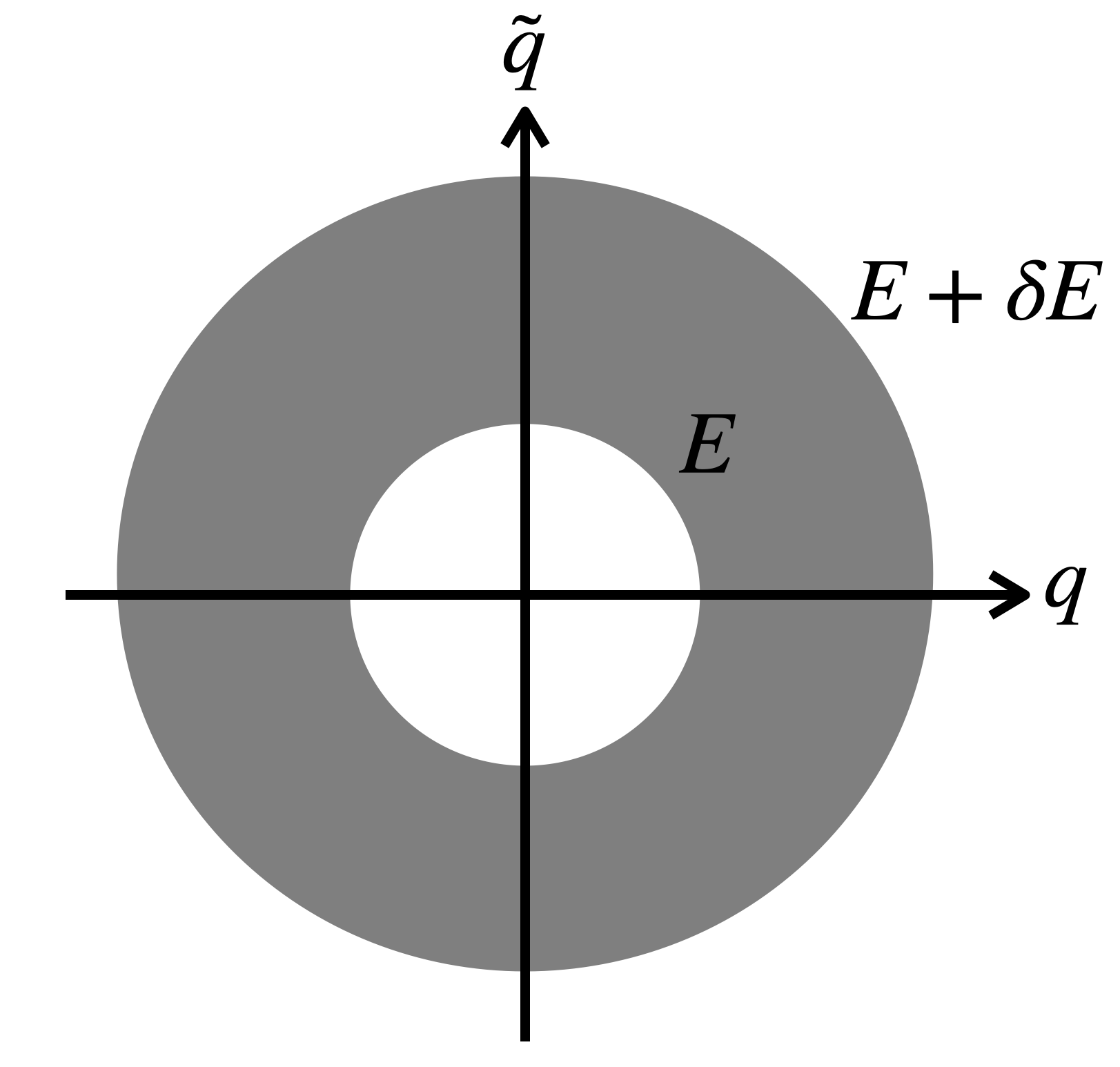}
    \caption{Ensemble in the energy range $E\rightarrow E+\delta E$ on the tangent bundle.}
    \label{QQ}
\end{figure}

In this section, we shall construct the statistical ensemble with the imaginary Lagrangian defined on the tangent bundle. We first would like to show that the area of the tangent bundle preserves under the time evolution. Suppose there is an element area: $d \Tilde{q}  d q\vert_{\tau=0}$ and a later time $\tau>0$ area is given by $d\Tilde{q} d q\vert_{\tau>0}$. Using the fact that $q_\tau=q_0+\Tilde{q} d\tau$ and $\Tilde{q} _\tau=\Tilde{q} _0+\Tilde{\Tilde{q}}  d\tau$, we have
\begin{eqnarray}
    d\Tilde{q} dq\vert_{\tau>0}&=&\left(d\Tilde{q}_0+d\Tilde{\Tilde{q}} d\tau\right)\left(dq_0+d\Tilde{q}  d\tau\right)\;.\nonumber\\
    &=&\left(d\Tilde{q}_0+\frac{\partial\Tilde{\Tilde{q}}}{\partial \Tilde{q}}d\Tilde{q} d\tau\right)\left(dq_0+\frac{\partial\Tilde{q} }{\partial q}dqd\tau\right)\nonumber\\
    &\approx & d \Tilde{q} d q\vert_{\tau=0}\left(1+\left(\frac{\partial\Tilde{\Tilde{q}}}{\partial \Tilde{q}}+\frac{\partial\Tilde{q}}{\partial  q}\right)d\tau\right)\;.\label{qq1}
\end{eqnarray}
Using \eqref{q11} and \eqref{q13}, equation \eqref{qq1} becomes
\begin{eqnarray}
d\Tilde{q} dq\vert_{\tau>0}&=&d \Tilde{q} d q\vert_{\tau=0}\left(1+\left(\frac{\partial}{\partial \Tilde{q}}\frac{\partial \mathcal L}{\partial  q}-\frac{\partial}{\partial  q}\frac{\partial \mathcal L}{\partial  \Tilde{q}}\right)d\tau\right)\nonumber\\
d\Tilde{q}  dq\vert_{\tau>0}&=&d \Tilde{q} d q\vert_{\tau=0}\;.\label{SSS}
\end{eqnarray}
Therefore, the area on the tangent bundle is preserved under the time evolution and we shall treat is feature as a Lagrangian version of the Liouville's theorem, see figure \ref{LT1}. 
\\
\\
It now prompts to define a density function of the states on the tangent bundle: $\rho(\Tilde{q},q;\tau)$ and it is not difficult to see that the density function satisfies the equation
\begin{equation}
    0=\frac{\partial\rho}{\partial \tau}+
    \frac{\partial \rho}{\partial q}\frac{\partial q}{\partial \tau}+\frac{\partial \rho}{\partial \Tilde{q}}\frac{\partial \Tilde{q}}{\partial \tau}=\frac{\partial\rho}{\partial \tau}+\{\rho,\mathcal L\}\;,
\end{equation}
where $\{*,\mathcal L\}$ is treated as an imaginary-time Lagrangian bracket
\begin{equation}
    \{*,\mathcal L\}=\left(\frac{\partial *}{\partial \Tilde{q}}\frac{\partial \mathcal L}{\partial q}-\frac{\partial *}{\partial q}\frac{\partial \mathcal L}{\partial \Tilde{q}}\right)\;\label{qb}
\end{equation}
with a property $\{*,\mathcal L\}=-\{\mathcal L,*\}$. Here, $*$ is a function defined on the tangent bundle. Interesting fact is that the Lagrange bracket \eqref{qb} provides a set of dynamic equations as follows
\begin{eqnarray}
    \Tilde{q}&=&-\frac{\partial \mathcal L}{\partial \Tilde{q}}\;,\label{KK}\\
    \Tilde{\Tilde{q}}&=&\frac{\partial \mathcal L}{\partial q}\;,\label{KK2}
\end{eqnarray}
which are \eqref{q11} and \eqref{q13}, respectively. One can note that a set of \eqref{KK} and \eqref{KK2} can be treated as a Lagrange's version of the Hamilton's equations.
\\
\\
Let $dN(\Tilde{q},q;\tau)$ be a number of points in an element area $d\Tilde{q}dq$ around a point $(\Tilde{q},q)$. Then probability density is given by
\begin{equation}
 \rho(\Tilde{q},q;\tau)d\Tilde{q}dq=\lim_{N\rightarrow\infty}\frac{dN}{N}\;,
\end{equation}
which satisfies the normalisation condition
\begin{equation}
    \int_\Gamma d\Gamma \rho (\Tilde{q},q;\tau)=1\;,\;\;\;d\Gamma=d\Tilde{q}dq\;.
\end{equation}
Any macroscopic quantity can be computed through 
\begin{equation}
    \langle \mathcal Q(\Tilde{q},q)\rangle=\int d\Gamma\rho(\Tilde{q},q;\tau)\mathcal Q(\Tilde{q},q)\;,
\end{equation}
which is treated as an ensemble average. Moreover, the time evolution of the ensemble average is given by
\begin{eqnarray}
    \frac{d}{d\tau}\langle \mathcal Q(\Tilde{q},q)\rangle=\int d\Gamma\frac{\partial\rho(\Tilde{q},q;\tau)}{\partial \tau}\mathcal Q(\Tilde{q},q)=\int d\Gamma \mathcal Q(\Tilde{q},q)\left(\frac{\partial \rho}{\partial \Tilde{q}}\frac{\partial \mathcal L}{\partial q}-\frac{\partial \rho}{\partial q}\frac{\partial \mathcal L}{\partial \Tilde{q}}\right)\;.
\end{eqnarray}
Applying integrating by parts, we obtain
\begin{eqnarray}
    \frac{d}{d\tau}\langle \mathcal Q(\Tilde{q},q)\rangle&=&-\int d\Gamma \rho \left[\left(\frac{\partial \mathcal Q}{\partial \Tilde{q}}\frac{\partial \mathcal L}{\partial q}-\frac{\partial \mathcal Q}{\partial q}\frac{\partial \mathcal L}{\partial \Tilde{q}}\right)+\mathcal{Q}\left(\frac{\partial}{\partial \Tilde{q}}\frac{\partial \mathcal L}{\partial q}-\frac{\partial}{\partial q}\frac{\partial \mathcal L}{\partial \Tilde{q}}\right)\right]\;.\nonumber\\
    &=&-\int d\Gamma \rho\{\mathcal L,\mathcal{Q} \}=\langle \{\mathcal Q,\mathcal L \} \rangle.
\end{eqnarray}
For an equilibrium macroscopic state, the ensemble average does not explicitly depend on time: $\frac{\partial\rho}{\partial\tau}=0$, demanding 
\begin{equation}
    \{\rho_{eq},\mathcal L\}=0\;.
\end{equation}
This means that a possible solution is for $\rho_{eq}$ to be a function of the Lagrangian: $\rho_{eq}(\mathcal L(\Tilde{q},q))$. Consequently, one finds that 
\begin{equation}
    \{\rho_{eq}(\mathcal L),\mathcal L \}=\rho_{eq}'(\mathcal L)\{\mathcal L,\mathcal L\}=0\;.
\end{equation}
This means that $\rho$ is constant on the energy surface $E=-\mathcal L$, in tangent bundle.
\\
\\
\textbf{Microcanonical ensemble}: For a system with fixed internal energy $U=\langle E\rangle$, volume $V$ and a number of particles $n$, the density function is given by
\begin{equation}
    \rho(\Tilde{q},q)=\frac{1}{\Sigma}\delta (\mathcal L(\Tilde{q},q)+U)\;.
\end{equation}
With the $\int \rho(\Tilde{q},q) d\Gamma=1$, it demands
\begin{equation}
    \Sigma(U)=\int d\Gamma \delta(\mathcal L(\Tilde{q},q)+U)\;.
\end{equation}
Then the ensemble defined on tangent bundle with energy less than or equal $U$ is given by
\begin{equation}
    \Omega(U)=\int_{-\mathcal L(\Tilde{q},q)\leq U} d\Gamma=\int_0^U dE \Sigma (E)\;.\label{En1}
\end{equation}
With a given ensemble in \eqref{En1}, the classical statistical entropy is given by
\begin{equation}
    S(U,V,n)=k_B\ln \Omega(U)\;.
\end{equation}
For two separated systems with $(U_1,V_1,n_1)$ and $(U_2,V_2,n_2)$, the total ensemble is given by $\Omega_{12}=\Omega_1\Omega_2$ resulting in
\begin{equation}
    S_{12}=S_1+S_2\;,\label{S12}
\end{equation}
which is known as the additive property of the entropy.
\\
\\
Now, we are ready to make a connection between microscopic and macroscopic worlds. Let's start with a definition of the temperature 
\begin{equation}
   \frac{1}{T}=\left(\frac{\partial S}{\partial U}\right)\Big\vert_{V,n}\;.
\end{equation}
\begin{equation}
    dS_{12}=dS_1+dS_2\;.
\end{equation}
At equilibrium, $dS_{12}=0$ and $d(U_1+U_2)=0$, one obtains
\begin{equation}
    \frac{1}{T_1}=\left(\frac{\partial S_1}{\partial U_1}\right)\Big\vert_{V_1,n_1}=\left(\frac{\partial S_2}{\partial U_2}\right)\Big\vert_{V_2,n_2}=\frac{1}{T_2}\;,
\end{equation}
which gives a thermal equilibrium condition between two systems or the zeroth law of thermodynamics. 
\\
\\
\textbf{Example}: We shall now consider the imaginary-time Lagrangian for the harmonic oscillator
\begin{equation}
    -\mathcal L(\Tilde{q},q;\tau)=\left(\Tilde{q}^2+q^2\right)=E\;.
\end{equation}
This new form of the Lagrangian allows us to construct the classical ensemble on tangent bundle.
In the energy range: $E\rightarrow E+\delta E$, see figure \ref{QQ}, the number of microstates is give by
\begin{equation}
    \delta \Omega=2\pi\delta E\;,
\end{equation}
and, obviously, in the energy range $0\rightarrow E$, the number of microstates is given by
\begin{equation}
     \Omega(E)=\frac{1}{h}\int_0^Ed\Omega =\frac{E}{\hbar} \;.
\end{equation}
The factor $h$ is introduced solely to render the phase-space volume dimensionless, in analogy with standard classical statistical mechanics, and does not imply quantization. Applying the relation $U=\langle E\rangle=k_BT$ and the first law $\delta Q=\delta U$, one finds
\begin{equation}
    \frac{\delta \Omega}{\Omega}=\frac{\delta U}{U}=\frac{\delta Q}{k_BT}\rightarrow dS=\frac{\delta Q}{T}\sim \frac{\delta \Omega}{\Omega}
\end{equation}
Finally, we find that the entropy is proportional to the total number of microstates: $S =k_B \ln \Omega(E)$.
\\
\\
\textbf{Canonical ensemble}: A system 1 characterised by $(U_1,V_1,n_1)$ is in thermal equilibrium at temperature $T$ with a heat bath, labeled as 2, characterised by $(E_2,V_2,n_2)$ with conditions $E_2 \gg E_1$ and $n_2 \gg n_1$ and $E_{12}=E_1+E_2$. The condition here is that the energy is allowed to exchange but not of particles. The Lagrangian of the total system is given by
\begin{equation}
    \mathcal L_{12}(\Tilde{q}_1,\Tilde{q}_2,q_1,q_2)=\mathcal L_1(\Tilde{q}_1,q_1)+\mathcal L_2(\Tilde{q}_2,q_2)\;.
\end{equation}
Since the total system is isolated, the density function of the total system is given by
\begin{equation}
    \rho_{12}(\Tilde{q},q)=\frac{1}{\Sigma_{12}}\delta(\mathcal L_{12}+E_{12})\;,
\end{equation}
where $\Tilde{q}=(\Tilde{q}_1,\Tilde{q}_2)$ and $q=(q_1,q_2)$ and 
\begin{equation}
    \Sigma_{12}=\int d\Gamma_{12} \delta(\mathcal L_{12}+E_{12})\;.
\end{equation}
Then the classical ensemble is given by
\begin{equation}
    \Omega_{12}(E_{12})=\int d\Gamma_{12} \delta(\mathcal L_{12}+E_{12})\;.
\end{equation}
Actually, we are interested in study the property the system 1. Then one has to trace out 2
\begin{eqnarray}\label{123}
    \rho_1(\Tilde{q}_1,q_1)&=&Tr_2 \rho_{12}(\Tilde{q},q)\nonumber\\
    &=&\frac{\int d\Tilde{q}_2\int dq_2\delta(\mathcal L_1+\mathcal L_2+E_{12})}{\Omega_{12}(E_{12})}\nonumber\\
    &=&\frac{\Omega_2(E_{12}+\mathcal L_1)}{\Omega_{12}(E_{12})}\;,
\end{eqnarray}
where $\Omega_2(E_2)=\Omega_{12}(E_{12}+\mathcal L_1)$ is a classical ensemble for the system 2. With a condition $E_1\ll E_{12}$, one can expand $\ln\Omega_2(E_2)$ around the $E_2=E_{12}$ resulting in 
\begin{equation}
    \ln \Omega_2(E_2)\approx \ln \Omega_2(E_{12})+\frac{\partial \ln\Omega_2}{\partial E_2}\Big\vert_{E_2=E_{12}}\mathcal L_1\;.
\end{equation}
Since the system 1 and system 2 are in thermal equilibrium with temperature $T$, then we have
\begin{equation}\label{r12}
\Omega_2(E_{12}+\mathcal L_1)=\Omega_2(E_{12})e^{\frac{\mathcal L_1}{K_BT}}
\end{equation}
Inserting \eqref{r12} into \eqref{123}, one gets
\begin{equation}
    \rho_1(\Tilde{q}_1,q_1)=\frac{\Omega_2(E_{12})}{\Omega_{12}(E_{12})}e^{\frac{\mathcal L_1}{K_BT}}=\frac{e^{\beta \mathcal L_1}}{\int d\Gamma_1e^{\beta \mathcal L_1}}\propto e^{\beta \mathcal L_1} \;,
\end{equation}
where $e^{\beta \mathcal L_1}$ will be treated as a Lagrangian version of the Boltzmann factor and $\beta=\frac{1}{k_BT}$.
\\
\\
What we have now is that, for any system with thermal equilibrium with surrounding, the density function is given by
\begin{equation}
    \rho(\Tilde{q},q)=\frac{e^{\beta \mathcal L(\Tilde{q},q)}}{\int d\Gamma e^{\beta \mathcal L(\Tilde{q},q)}}\;,
\end{equation}
and the canonical partition function $Z$ is defined as
\begin{equation}
    Z\equiv-\frac{1}{h}\int d\Gamma e^{\beta \mathcal L(\Tilde{q},q)}\;.
\end{equation}
Next, we consider
\begin{eqnarray}
    \frac{\partial}{\partial \beta} \ln Z&=&-\frac{\int d\Gamma (\mathcal L)e^{\beta L}}{\int d\Gamma e^{\beta \mathcal L}}\nonumber\\
    &=&\langle -\mathcal L\rangle =U\;.
\end{eqnarray}
Then we employ the relation $U=\frac{\partial}{\partial \beta}(\beta F)$, where $F$ is the Helmholtz free energy. The final relation is 
\begin{equation}
    Z=e^{\beta F}\;,\;\;\text{or}\;\; F(T,V,n)=k_BT\ln Z(T)\;.
\end{equation}
\textbf{Example}: We shall work out with the harmonic oscillator with one degree of freedom again. The partition is given by
\begin{equation}
    Z=-\frac{1}{h}\int_{-\infty}^{+\infty}d\Tilde{q}\int_{-\infty}^{+\infty}dq e^{\beta(\Tilde{q}^2+q^2)}=\frac{k_BT}{\hbar}\;,
\end{equation}
which is identical with the Hamiltonian approach.

\section{Concluding summary}\label{SS}
We have presented a systematic construction of classical statistical ensembles formulated on the tangent bundle $(i\Tilde{q},q)$. A key technical step in this approach is the Wick rotation from real time to imaginary time applied to the Lagrangian. This transformation leads naturally to a set of first-order differential equations, which can be interpreted as the Lagrangian counterpart of Hamilton’s equations written in velocity variables. Within this imaginary-time framework, the tangent bundle acquires a well-defined geometric structure under which the area element is preserved along the dynamical flow. This property reflects the Hamiltonian nature of the induced dynamics and ensures the existence of an invariant measure, playing a role analogous to Liouville’s theorem in the present formulation. As a result, classical statistical ensembles can be constructed directly on the tangent bundle, which is endowed with the symplectic structure $d\Tilde{q}\wedge dq$. The corresponding statistical (Boltzmann) entropy then follows in a natural manner. Using the one-dimensional harmonic oscillator as an explicit example, we have shown that this imaginary-time Lagrangian formulation reproduces exactly the same thermodynamic quantities, such as the entropy and canonical partition function, as those obtained from the standard Hamiltonian approach. This agreement confirms the internal consistency of the framework for systems with quadratic kinetic terms\footnote{In this work, we restrict our attention to non-relativistic systems.}.
\\
\\
We hope that this preliminary study fills the missing element illustrated in Fig.~\ref{F1}, thereby completing the overall picture connecting Lagrangian and Hamiltonian formulations of classical statistical mechanics. Moreover, this alternative Lagrangian-based perspective may open new mathematical directions for exploring the geometric structure of statistical mechanics and related physical systems.

\begin{acknowledgements}

The author would like to acknowledge the use of ChatGPT, a large language model developed by OpenAI, for assistance in improving the clarity of presentation, refining the exposition, and identifying potential conceptual and stylistic issues during the revision of this manuscript. The author takes full responsibility for the content, interpretation, and conclusions presented in this work.
\end{acknowledgements}

\end{document}